\documentclass[letter]{aa} 

\bibpunct{(}{)}{;}{a}{}{,}
\usepackage{graphicx}
\usepackage{txfonts}
\begin{document}

   \title{Widespread Occurrence of High-Velocity Upflows in Solar Active Regions}

   \author{S. L. Yardley
          \inst{1}
          \and
          D. H. Brooks\inst{2}\fnmsep\inst{3}
          \and
          D. Baker\inst{1}
          }

\institute{University College London, Mullard Space Science Laboratory, Holmbury St. Mary, Dorking, Surrey, RH5 6NT, UK \\
              \email{stephanie.yardley@ucl.ac.uk}
         \and
             College of Science, George Mason University, 4400 University Drive, Fairfax, VA 22030 USA\\
             \email{dhbrooks.work@gmail.com}
        \and
            Hinode Team, ISAS/JAXA, 3-1-1 Yoshinodai, Chuo-ku, Sagamihara, Kanagawa 252-5210, Japan
            }
             
   \date{Received 19 April 2021}

 
  \abstract{}{We performed a systematic study of 12 active regions (ARs) with a broad range of areas, magnetic flux and associated solar activity in order to determine whether there are upflows present at the AR boundaries and if these upflows exist, whether there is a high speed asymmetric blue wing component present in the upflows.}
   {To identify the presence and locations of the AR upflows we derive relative Doppler velocity maps by fitting a Gaussian function to {\it Hinode}/EIS Fe XII 192.394\,\AA\ line profiles. To determine whether there is a high speed asymmetric component present in the AR upflows we fit a double Gaussian function to the Fe XII 192.394\,\AA\ mean spectrum that is computed in a region of interest situated in the AR upflows.}
   {Upflows are observed at both the east and west boundaries of all ARs in our sample with average upflow velocities ranging between -5 to -26~km s$^{-1}$. A blue wing asymmetry is present in every line profile. The intensity ratio between the minor high speed asymmetric Gaussian component compared to the main component is relatively small for the majority of regions however, in a minority of cases (8/30) the ratios are large and range between 20 to 56~\%.}
   {These results suggest that upflows and the high speed asymmetric blue wing component are a common feature of all ARs.}

            
    \titlerunning{Asymmetries in AR upflows}
    \authorrunning{Yardley et al.}

   \maketitle

%

\section{Introduction}
Identifying the origin of the slow speed ($\sim$500\,km s$^{-1}$) solar wind and determining the structure and dynamics at its sources are amongst the most important goals of current research in heliophysics. With the recent launch of {\it Parker Solar Probe} \citep{Muller-2020} and {\it Solar Orbiter} \citep{Fox-2016} there is great interest in connecting observations made {\it in-situ} in the solar wind with remote sensing measurements of the source regions. As part of that effort, characterizing the physical properties of potential source regions is important.

Previous observations from the {\it Hinode} satellite \citep{Kosugi2007} have shown that high temperature upflows are present at the boundaries of many solar active regions \citep{Sakao2007,Harra2008,DelZanna2008,Tian2021}. These upflows may be outflows if they have access to open magnetic fields that connect to the heliosphere \citep{Sakao2007,Harra2008,Doschek2008, Baker2009} and elemental abundance measurements show that they have a coronal composition \citep{Brooks2011}. More precisely, the upflowing material may be a source of the slow solar wind. 
\citet{Brooks2015} found, by using full-sun composition and Doppler maps along with the potential-field source-surface (PFSS) model, that the primary source contributing to the slow solar wind does appear to originate from the boundaries of active regions. Conversely, by using a global potential field model, \citet{Edwards2016} determined that active region upflows were not associated with open magnetic fields. However, the study by \citet{Edwards2016} only concentrated on analysing upflows present in a limited number of active regions. Furthermore, plasma that is confined along closed magnetic field could also contribute to the slow solar wind indirectly if it can escape along open magnetic field through reconnection. The \citet{Hinode2019} describe the current state of active region outflow studies within the wider context of solar wind research. For a more general review of solar wind studies see, e.g., \citet{Abbo2016}.     

Typical bulk flow speeds measured in the upflows by the EUV Imaging Spectrometer \citep[EIS,][]{Culhane2007} are 10--40\,km s$^{-1}$ with mass motions of 30--60\,km s$^{-1}$ \citep{Brooks2011}, but there is often a high speed asymmetric component in the blue wing of the EIS spectral line profiles. This high speed component is weak, with the intensity only amounting to around 10\% of the total emission of the upflow \citep{Brooks2012}. This letter focuses on key questions about these upflows and asymmetries.

Browsing the EIS database leaves the impression that the  upflows are present in most active regions, but selection effects may play a role in this. For example, {\it Hinode} generally observes the largest, most active targets, and there is often a delay between the emergence of a new active region, and its selection for observations. Furthermore, since observing the emergence phase is difficult, the formation of the upflows is often missed. If this is a quick process, it might well be the case that upflows exist for most of the lifetime of an active region \citep{Demoulin2013,Zangrilli2016,Baker2017,Harra2017}. In contrast, we know that the evolution of the magnetic field  affects how an active region decays and finally disperses \citep{Baker2015,vandriel2015}, so at some point this process will affect the magnetic field at the active region boundary and potentially disrupt the upflows. 

The high speed component of the EIS line profiles has also received much attention but has only been observed in a relatively small number of case studies. Asymmetries are seen in and around the footpoints of coronal loops \citep{Hara2008,Bryans2010,Tian2021}, and have also been linked with chromospheric jets \citep{DePontieu2009}. The magnitude of the asymmetries appears to depend on temperature and they are often more pronounced in the upflows \citep{Brooks2012}, but many of these results are known from the analysis of single active regions. Since the asymmetries are weak and measurements are difficult there may also be a tendency to analyze examples with a strong signal. This could potentially be misleading.

Here we perform the first systematic study of 12 active regions, three of which are observed for a second time later in their evolution, in order to answer two questions:
(i) are upflows present in all sizes of active regions, and (ii) if upflows are present, is there always a high speed blue wing asymmetry
in the high temperature line profiles? 

The 12 active regions we study cover a range of activity levels and conditions. They have previously been analyzed for different purposes by \citet{Warren2012}, \citet{Brooks2016}, and \citet{Viall2017}, so we already know a great deal about them. The total unsigned magnetic flux spans a factor of $\sim$7 (4.08$\times$10$^{21}$--2.73$\times$10$^{22}$\,Mx), and the active region area varies by a factor of $\sim$5 (2.87$\times$10$^{19}$--1.48$\times$10$^{20}$\,cm$^2$). The emission measure distributions in the cores of the active regions are generally strongly peaked near 4\,MK, with a tendency to be steeper for the regions of stronger magnetic flux. Mass motions in the active region core loops are typically modest ($\sim$17\,km s$^{-1}$). There is evidence of widespread cooling from high temperatures in all these regions \citep{Viall2017}. Only one of the active regions is in close proximity to a coronal hole therefore, there is little possibility of interactions between the active region boundaries and neighbouring coronal holes. PFSS extrapolations of the regions, however, show that there is open magnetic field present at one or both of the active region boundaries for seven out of twelve active regions. These regions of open field are located close to the areas of upflow we analyze. This suggests that for these regions the upflows are indeed outflows however, this requires further investigation.

\section{Observations and Methods} \label{sec:obs}

Table~\ref{Table1} gives information on the NOAA ARs and their disk location, along with the EIS observations we used for this analysis.
The EIS data were processed using the standard eis\_prep procedure available in {\it SolarSoft}. This routine accounts for the CCD
dark current and pedestal, and handles dusty, hot, and warm pixels and cosmic rays. For this study, however, we did not apply the photometric calibration.
Previous work has shown that applying the calibration modifies the spectral line width \citep{Brooks2016,Testa2016}, and since this is a
crucial parameter for studying the shape of the line profile we decided to use the raw uncalibrated spectra.

EIS observes two wavelength bands from 171--212\,\AA\, and 245--291\,\AA. The spectral resolution of the instrument is 22\,m\AA. The datasets listed in Table~\ref{Table1} were obtained in a variety of observing modes using both the 1$''$ and 2$''$ slits and exposure times of 30--60\,s. A field-of-view of 100--360$''$ was scanned in the solar-X direction while a slit height (in solar-Y) of 240--512$''$ was used. Specific details for each dataset were provided by \citet{Brooks2016}.

We use the Fe XII 192.394\,\AA\, spectral line for this analysis. This is the second strongest, clean and unblended line in EIS active region spectra \citep{Young2007,Brown2008} and is present in all our datasets. The line is formed at 1.6\,MK\, in equilibrium conditions, and this is a good representative temperature for measuring upflows based on previous studies. To derive relative Doppler velocities from the Fe XII 192.394\,\AA\, line we first remove the orbital drift of the EIS spectrum on the CCD using the neural network model of \citet{Kamio2010}. This method uses instrument temperature information to model the drift of the Fe XII 195.119\,\AA\, line as the satellite moves around its orbit. Since the Fe XII 192.394\,\AA\, and Fe XII 195.119\,\AA\, lines are very close in wavelength the orbital correction for one should be valid for the other. In some cases we also apply an additional correction to account for any residual orbital variation that was not removed \citep[see, e.g.,][]{Brooks2020b}.
To create relative Doppler velocity maps we fit a Gaussian function to the Fe XII 192.394\,\AA\, spectrum across each of the scans.
These maps allow us to easily identify upflow locations and determine whether they exist in every AR.

There are multiple methods that can be used to analyse the high-speed asymmetric component such as a double Gaussian fit \citep{Peter2010, Bryans2010, Brooks2012}, blue-red (B-R) asymmetry analysis \citep{DePontieu2009, Tian2011}, and the B-R guided double Gaussian fit \citep{DePontieu2010, Tian2011} . In this study, to assess whether a high-speed asymmetric component is present in the Fe XII 192.394\,\AA\, line profile, we use a double Gaussian function. We first select a region of interest in the upflow and calculate the mean spectrum. We then fit a single Gaussian function and use that to construct a double Gaussian template. The minor high speed component is assumed to have 10\% of the intensity of the main component, and both Gaussians are assumed to have the same width. The centroid of the minor component is initially placed in the blue wing with a Doppler shift of 125\,km s$^{-1}$. This value is based on the results from \citet{Brooks2012}; of course the fit itself determines the actual Doppler shift. The assumption that the
two components have the same width may not perfectly capture the smooth shape of the asymmetry, but it provides a consistent way of quantifying e.g. the area of the minor component compared to the main component across all the different ARs.

\begin{table*}
\caption{The NOAA ARs and EIS data used in this study. The information in this table has been taken from \citet{Warren2012}. In columns 4 and 5, X$_{c}$ and $Y_{c}$ correspond to the X and Y coordinates of the active region that have been differentially rotated to the center of the EIS raster.}      
\label{Table1}    
\centering
\begin{tabular}{c c c c c c}
\hline
No. & NOAA & Time & X$_{c}$& Y$_{c}$ & EIS File\\
& AR & (UT) & (arcsec) & (arcsec) & \\
\hline 
1 & 11082 & 2010 Jun 19 01:57:44 & -306.4 & 439.3 & eis\_l1\_20100619\_014433 \\
2 & 11082 & 2010 Jun 21 01:46:37 & 162.9 & 405.2 & eis\_l1\_20100621\_011541 \\
3 & 11089 & 2010 Jul 23 15:03:07 & -363.4 & -453.6 & eis\_l1\_20100723\_143210 \\ 
4 & 11109 & 2010 Sep 29 23:51:36 & 361.5 & 261.5 & eis\_l1\_20100929\_223226 \\
5 & 11147 & 2011 Jan 21 14:10:50 & 26.6 & 476.5 & eis\_l1\_20110121\_133954\\
6 & 11150 & 2011 Jan 31 11:25:19 & -470.0 & -250.6 & eis\_l1\_20110131\_102326 \\
7 & 11158 & 2011 Feb 12 15:32:13 & -248.4 & -211.8 & eis\_l1\_20110212\_143019\\
8 & 11190 & 2011 Apr 11 12:00:42 & -492.6 & 281.0 & eis\_l1\_20110411\_105848 \\
9 & 11190 & 2011 Apr 15 01:17:19 & 218.1 & 304.4 & eis\_l1\_20110415\_001526 \\
10 & 11193 & 2011 Apr 19 13:32:20 & 36.3 & 363.5 & eis\_l1\_20110419\_123027 \\
11 & 11243 & 2011 Jul 2 03:38:08 & -299.0 & 216.6 & eis\_l1\_20110702\_030712 \\
12 & 11259 & 2011 Jul 25 09:36:09 & 224.7 & 323.4 & eis\_l1\_20110725\_090513 \\
13 & 11271 & 2011 Aug 21 12:25:42 & -50.8 & 150.8 & eis\_l1\_20110821\_105251 \\
14 & 11339 & 2011 Nov 8 19:14:27 & 88.1 & 258.4 & eis\_l1\_20111108\_181234 \\
15 & 11339 & 2011 Nov 10 11:33:19 & 406.0 & 266.8 & eis\_l1\_20111110\_100028 \\
\hline
\end{tabular}
\end{table*}

\section{Results} \label{sec:results}

\begin{figure*}[h]
\centering
\includegraphics[width=1.0\textwidth]{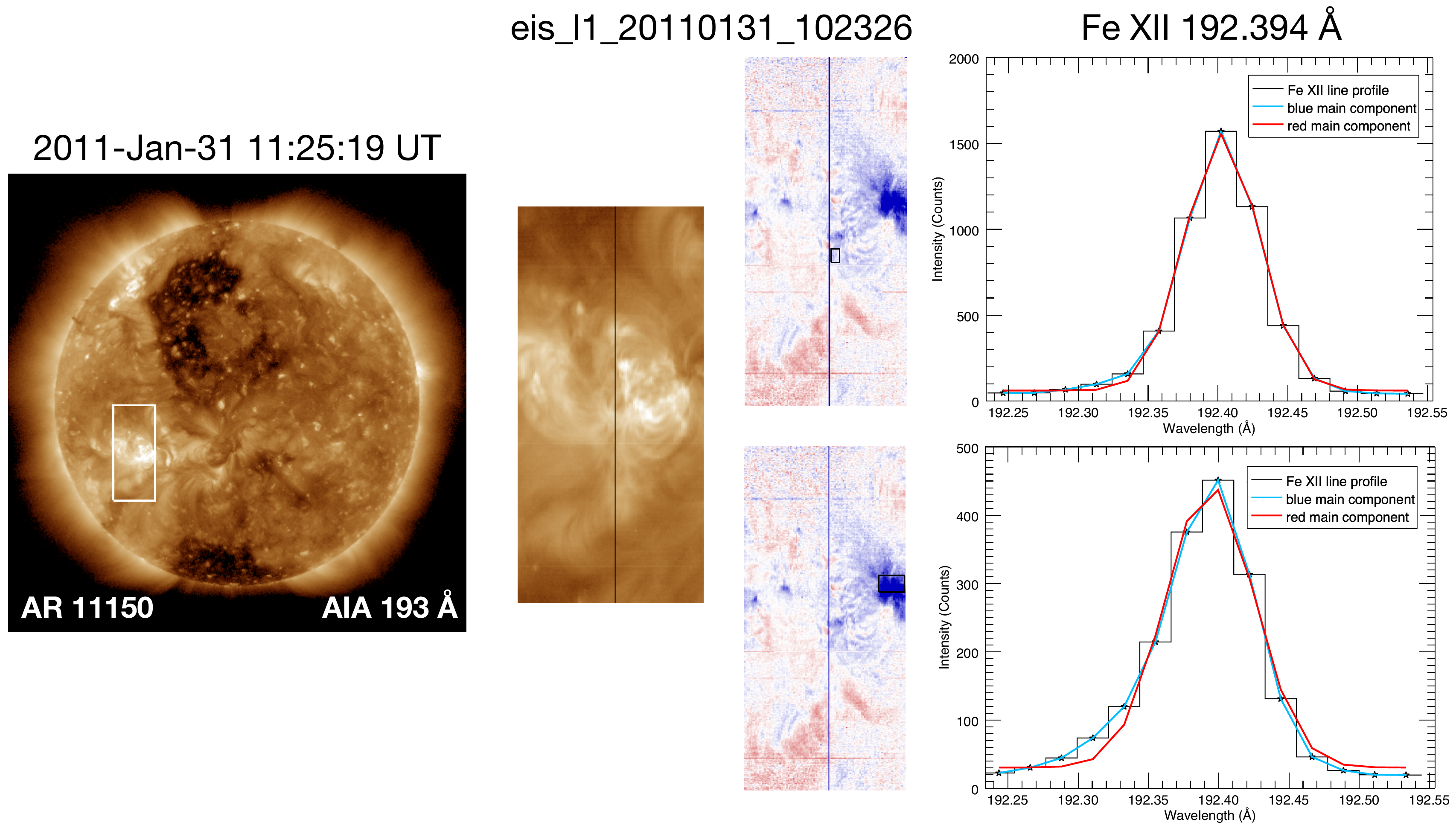}
\caption{Blue wing asymmetries for the east and west upflows of NOAA AR 11150 (no. 6 in Table~\ref{Table1}). The left panel shows the {\it SDO}/AIA 193\,\AA\ image with the white box corresponding to the EIS field-of-view. The middle panel shows the EIS Fe XII 192.394\,\AA\ image and Doppler velocity maps. The final panel shows the mean spectrum of the pixels highlighted by the black boxes in the EIS scans (middle panel) fitted with a single Gaussian. \label{fig:fig1} }
\end{figure*}

\begin{figure*}[h]
\centering
\includegraphics[width=1.0\textwidth]{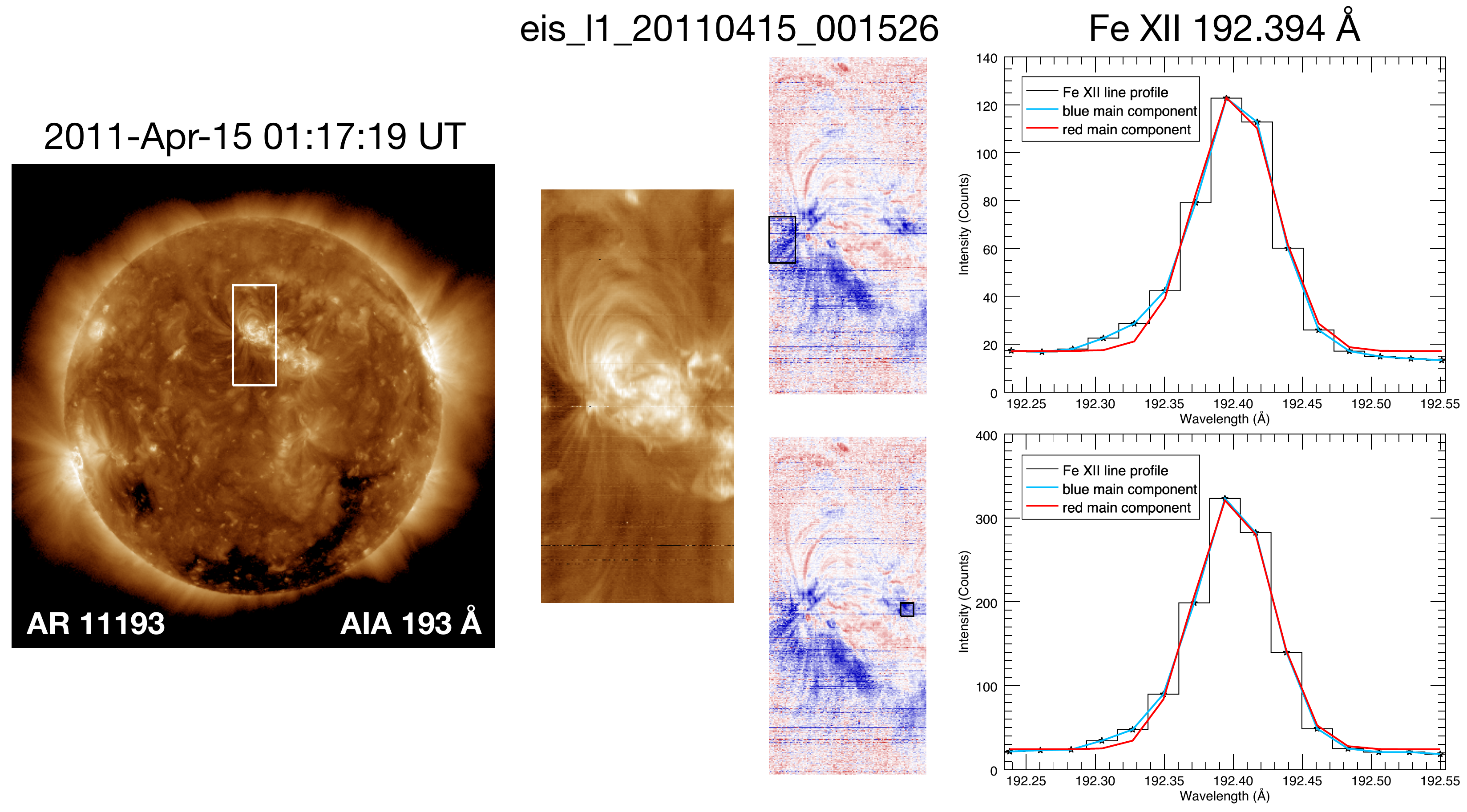}
\caption{Same as Figure~\ref{fig:fig1} but for NOAA AR 11193 (no. 10 in Table~\ref{Table1}). \label{fig:fig2} }
\end{figure*}

Figures~\ref{fig:fig1} and \ref{fig:fig2} show examples of the AR upflows and the asymmetric blue wing component present in the EIS Fe XII 192.394\,\AA\ spectral line for ARs 11150 and 11193 (no. 6 and 10 in Table~\ref{Table1}). The middle panels of both Figures show the EIS Fe XII image along with the Doppler velocity maps of the ARs that have been derived using the methods outlined in Section~\ref{sec:obs}. The Doppler velocity maps show that there are upflows (blue shifts) present at the east and west boundaries of both ARs. The upflows are particularly strong at the solar west boundary of AR 11150 whereas, the strongest upflow for AR 11193 is located at the east boundary. We found that these upflows were present at the east and west boundaries in the Doppler maps for each AR in our sample.

The mean spectral profile is given in the final panels of Figures~\ref{fig:fig1} and \ref{fig:fig2} for the pixels indicated by the black boxes in the Doppler maps (middle panel). Each spectral profile shows a blue wing asymmetry. This high speed asymmetric component is relatively weak although, its strength varies depending upon the upflow and selected region. We found that these blue wing asymmetries were present in all of the line profiles constructed from the regions of interest located in the AR upflows.

\begin{figure}[h]
\centering
\includegraphics[width=0.5\textwidth]{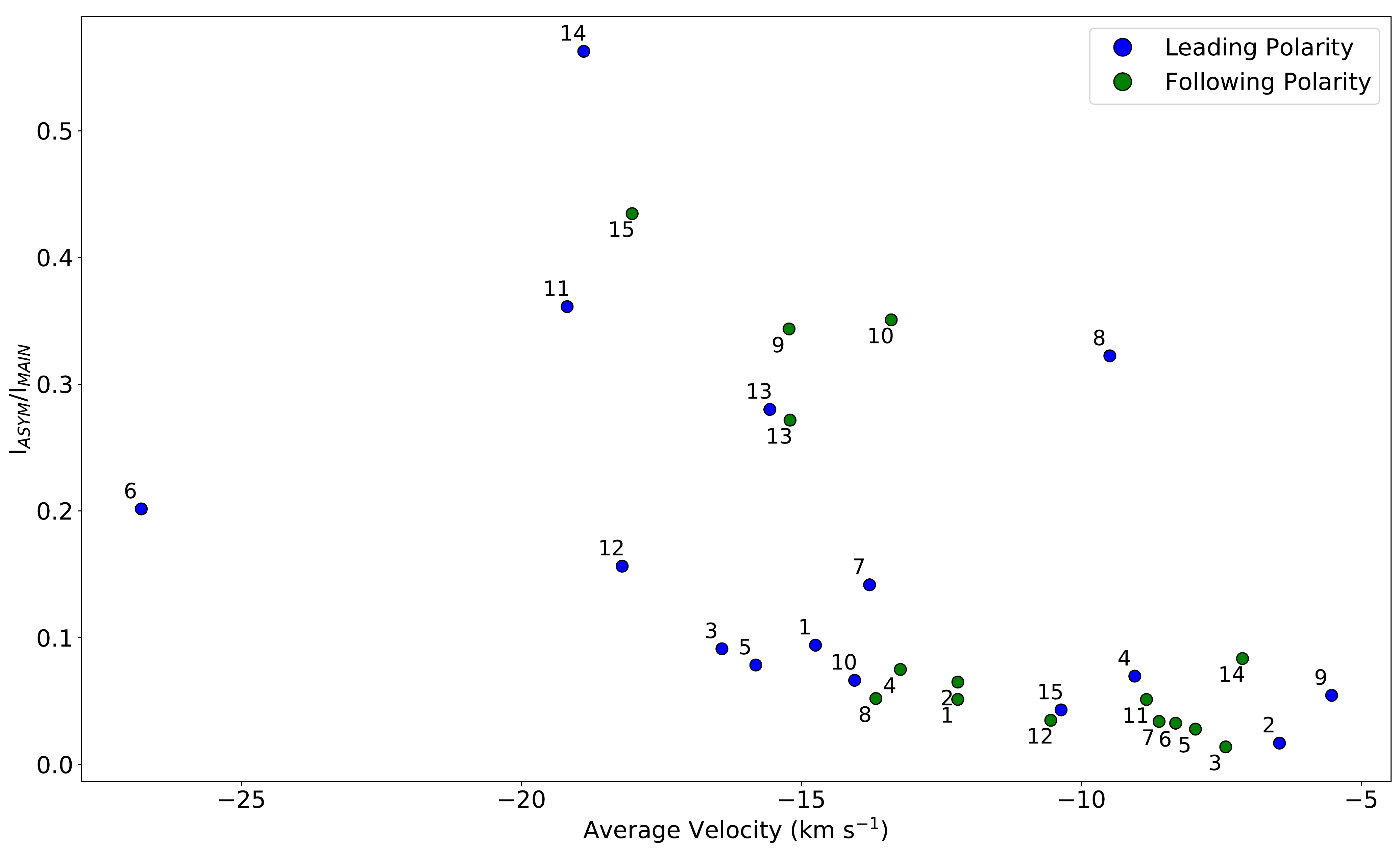}
\caption{The intensity ratio of the asymmetric and main component Gaussians as a function of the average velocities calculated for the regions of interest located in the east and west upflows of each AR. The blue (green) circles represent the average velocities associated with the leading (following) polarity of the AR. The numbers correspond to the EIS scans of the ARs given in Table~\ref{Table1}.  \label{fig:fig3} }
\end{figure}

Figure~\ref{fig:fig3} shows the intensity ratio of the asymmetric and main component Gaussians as a function of the average velocity calculated for the regions of interest for the east and west upflows for each AR. The average velocities have been shifted by -5~km~s$^{-1}$ to account for the formation temperature of the Fe XII line (see Figure 6 of \citealt{Peter1999}). 

The blue (green) points and corresponding labels represent the average velocities obtained for the leading (following) polarities of the AR observations given in Table~\ref{Table1}. The actual and true errors in the average velocities are found to be 
$\pm$~4.63 km~s$^{-1}$ and $\pm$~4.88~km~s$^{-1}$ and are calculated by taking the standard deviation of the average velocities with and without the additional residual orbital correction, discussed in Section \ref{sec:obs}, respectively. Figure~\ref{fig:fig3} shows that the average velocities for the region of interest in each AR upflow are negative and the blue wing asymmetric component is present in all ARs. There is  a moderately strong negative correlation (Pearson correlation coefficient of -0.58) between the intensity ratio of the minor asymmetric compared to the main component and the associated average velocity, though these quantities are not always independent. The average velocities for the AR upflows were found to be in the range of -5 to -26~km~s$^{-1}$. For the majority of the upflows the intensity of the minor high speed component is relatively small i.e. less than 20\% of the main component, however, for eight of the upflows the intensity ratio ranges from 20 to 56\%. In these cases the asymmetric component is rather broad due to the assumptions that have been made regarding the double Gaussian template. We have also found a moderate positive correlation between the total unsigned flux of the active regions and the intensity ratio (Pearson correlation coefficient of 0.49). The relative intensity varies widely depending upon the location of the upflow i.e. whether we are considering the upflow situated at the boundary of the leading or following polarity and the stage of evolution of the active region. Similarly, these variations are also observed for the intensity ratio as a function of average velocity, as shown in Figure~\ref{fig:fig3}. Finally, no correlation was found between the total unsigned magnetic flux and the average upflow velocity.

\section{Summary \& Discussion} \label{sec:summary}

We carried out a systematic study using the EIS observations of 12 active regions in order to determine: (i) whether upflows exist in all sizes of active regions and if so (ii) is there always a high speed blue wing asymmetry present in the Fe XII 192.394\,\AA\ line profile. The ARs in our sample spanned a wide range of total unsigned magnetic flux (4.08 $\times$ 10$^{21}$ -- 2.73 $\times$10$^{22}$~Mx) and areas (2.87 $\times$10$^{19}$ -- 1.48 $\times$ 10$^{21}$ cm$^{2})$. Three of our ARs were observed for a second time roughly 2 to 4 days after the initial EIS observations, giving a sample of 15 AR measurements in our study. By creating Doppler velocity maps from the EIS scans of each AR it was possible to identify whether upflows were present in each AR and their locations. We found that these upflows were present at the east and west boundaries of all ARs in our sample. This suggests that upflows at the boundary are a common feature of ARs irrespective of their size. The average velocities for the regions of interest located in the AR upflows were found to be between -5 and -26~km~s$^{-1}$.

\citet{Baker2017} also reported large-scale upflows observed at the peripheries of both polarities in a sample of 10 ARs with similar flux range to the ARs in this work. They found that the majority of the long-term evolution of the upflows is attributed to projection effects that are dependent upon the AR’s location on the disk. This effect is stronger in ARs associated with higher velocities. In this study, we have analysed EIS scans of ARs that were taken at numerous longitudes ranging between -492.6$''$ and 406.0$''$ and so projection effects will be present. In fact, there is a significant change in the upflows and intensity of the blue wing asymmetries after approximately 2 to 4 days for three ARs (11082, 11190, 11339 in Table~\ref{Table1}) that were observed by EIS more than once. However, in this particular study, we are only interested in whether these upflows and asymmetries are present at the peripheries of ARs and not their evolution.

To determine whether a high speed asymmetric component exists in the Fe XII 192.394\,\AA\ spectral line we used a double Gaussian function, which was based upon the mean spectrum calculated for a region of interest located in the upflows. We then measured the intensity of the minor asymmetric component to the main component for all ARs. While, we acknowledge that in some cases, by using a double Gaussian fit with two components, the width does not entirely capture the shape of the blue wing asymmetry it provides a way to consistently compare the area of the minor and main components. Furthermore, comparisons with artificial line profiles support
this analysis strategy \citep{Tian2011}. An alternative method would be to use an asymmetric Gaussian fit where the degree of asymmetry comes from the difference in the widths between the Gaussians that fit the red and blue wings, or to follow the (B-R) asymmetry analysis introduced by \citet{DePontieu2009}. Differences in methodology however, do not change our fundamental results because the blue wing asymmetries were discovered to be present in each AR. This indicates that these asymmetries are likely to be present in all ARs. For most of our ARs the intensity of the high speed component compared to the main component is quite small with an intensity ratio of up to 20\%. Although, for eight upflows the ratio of intensities are between 20 to 56\%. 

We also found indications of moderate negative (positive) correlations between the intensity ratio of the minor asymmetric and main components, and the average upflow velocities (total unsigned magnetic flux). However, the upflow velocity can vary significantly depending upon the location of the upflow therefore, projection effects are potentially important. This requires full EIS coverage of the active regions. Other considerations need to be taken into account such as the size and locations of the selected regions of interest and the fact that sizeable asymmetric components will shift a single Gaussian to the blue wing and will give larger Doppler velocities.

Prior observations have suggested that the upflows located at the boundaries of ARs, and the high speed component, are likely to be a source of the slow solar wind. The fact that they are present in all ARs suggests that their contribution could be significant. Previous studies have tried to link these features to {\it in-situ} measurements of the solar wind, and currently, we are investigating their role in the production of solar energetic particles (SEPs). Plasma composition measurements are important for SEP connection studies \citep{Brooks2021} and further work on this topic will be the focus of missions such as {\it Solar Orbiter} and {\it Parker Solar Probe}. In fact, \citet{Harra2021} have already analysed a series of small type III bursts observed by {\it Parker Solar Probe} in order to identify the source of SEPs. They found that the most likely source of the radio storm was an extended blue-shifted outflow region located at the eastern boundary of an expanding active region.

\begin{acknowledgements}
SLY would like to thank NERC for funding via the SWIMMR Aviation Risk Modelling (SWARM) Project, grant number NE/V002899/1. 
The work of DHB was performed under contract to the Naval Research Laboratory and was funded by the NASA Hinode program. DB is funded under STFC consolidated grant number ST/S000240/1.
Hinode is a Japanese mission developed and launched by ISAS/JAXA, with NAOJ as domestic partner and NASA and STFC (UK) as international partners. It is operated by these agencies in co-operation with ESA and NSC (Norway). 

\end{acknowledgements}

\bibliographystyle{aa} 
   \bibliography{aa}

\end{document}